\documentclass[%
 preprint,
 superscriptaddress,
longbibliography,
 amsmath,amssymb,
 aps,
prb,
]{revtex4-1}

\usepackage{ulem}
\usepackage{floatrow} 
\usepackage{amsmath}
\usepackage{amsfonts}
\usepackage{graphicx}
\usepackage{dcolumn}
\usepackage{multirow}
\usepackage{bm}
\usepackage{lipsum}
\usepackage{hyperref}
\usepackage{threeparttable}

\usepackage{tabularx}
\usepackage{soul}

\usepackage{xcolor} 
\definecolor{lightgray}{gray}{0.6}
\definecolor{medgray}{gray}{0.4}

\newif\ifptitle
\newif\ifpnumber
\newcounter{para}
\newcommand\ptitle[1]{\par\refstepcounter{para}
{\ifpnumber{\noindent\textcolor{lightgray}{\textbf{\thepara}}\indent}\fi}
{\ifptitle{\textcolor{lightgray}{\textbf{[{#1}]}}}\fi}}

\begin{document}
\newcounter{td}
\newcommand{\todo}[1]{
    \stepcounter{td}
    \textcolor{red}{{\arabic{td}.} #1}
}
\setstcolor{black}
\newcommand{\delete}[1]{
    \textcolor{red}{\st{#1}}
}
\newcommand{\add}[1]{
    \textcolor{red}{#1}
}

\title{Uncertainty-Aware Molecular Dynamics from Bayesian Active Learning for Phase Transformations and Thermal Transport in SiC}

\author{Yu Xie\footnote{xiey@g.harvard.edu}}
\affiliation{John A. Paulson School of Engineering and Applied Sciences, Harvard University}
 
\author{Jonathan Vandermause}
\affiliation{John A. Paulson School of Engineering and Applied
Sciences, Harvard University}
\affiliation{Department of Physics, Harvard University}

\author{Senja Ramakers}
\affiliation{Corporate Sector Research and Advance Engineering, Robert Bosch GmbH, Renningen}
\affiliation{Interdisciplinary Centre for Advanced Materials Simulation, Ruhr-Universit\"at Bochum}

\author{Nakib H. Protik}
\affiliation{John A. Paulson School of Engineering and Applied Sciences, Harvard University}

\author{Anders Johansson}
\affiliation{John A. Paulson School of Engineering and Applied Sciences, Harvard University}

\author{Boris Kozinsky\footnote{bkoz@seas.harvard.edu}}
\affiliation{John A. Paulson School of Engineering and Applied Sciences, Harvard University}
\affiliation{Robert Bosch LLC, Research and Technology Center}



\begin{abstract}
Machine learning interatomic force fields are promising for combining high computational efficiency and accuracy in modeling quantum interactions and simulating atomistic dynamics. Active learning methods have been recently developed to train force fields efficiently and automatically. 
Among them, Bayesian active learning utilizes principled uncertainty quantification to make data acquisition decisions. In this work, we present a general Bayesian active learning workflow, where the force field is constructed from a sparse Gaussian process regression model based on atomic cluster expansion descriptors. To circumvent the high computational cost of the sparse Gaussian process uncertainty calculation, we formulate a high-performance approximate mapping of the uncertainty and demonstrate a speedup of several orders of magnitude.
 We demonstrate the autonomous active learning workflow by training a Bayesian force field model for silicon carbide (SiC) polymorphs in only a few days of computer time and show that pressure-induced phase transformations are accurately captured. The resulting model exhibits close agreement with both \textit{ab initio} calculations and experimental measurements, and outperforms existing empirical models on vibrational and thermal properties.
The active learning workflow readily generalizes to a wide range of material systems and accelerates their computational understanding.
\end{abstract}

\maketitle


\section{\label{sec:intro}Introduction}


\ptitle{The justification for these objectives: Why is the work important?}
Machine learning interatomic force fields have recently emerged as powerful tools in modeling interatomic interactions. They are capable of reaching near-quantum accuracy while being orders of magnitude faster than \textit{ab initio} methods\cite{behler2007generalized,shapeev2016moment,osti_1183138,schutt2017schnet,bartok2010gaussian,vandermause2020fly,batzner20223,chen2019graph,chen2022universal}.

Recently, efficient active learning schemes have been demonstrated for high-efficiency data collection, where molecular dynamics (MD) is driven by the machine learning force field, and only configurations satisfying certain acquisition criteria \cite{jinnouchi2019phase, jinnouchi2019fly, li2015molecular, podryabinkin2017active, vandermause2020fly, hodapp2020operando, young2021transferable} are computed with accurate but expensive DFT calculations and added to the training set. 
Among them, the FLARE \cite{vandermause2020fly} framework utilizes principled Gaussian process (GP) uncertainties to construct a Bayesian force field (BFF), a force field equipped with internal uncertainty quantification from Bayesian inference, enabling a fully autonomous active learning workflow.

\ptitle{Here insert sentences about what was missing in the previous work. What challenges were there? What is new here?}
The cost of prediction of conventional GP models scales linearly with the training set size, making it computationally expensive for large data sets. 
The sparse Gaussian process (SGP) approach selects a set of representative atomic environments from the entire training set to build an approximate model, which can scale to a larger training data set, but still suffers from the linear scaling of the inference cost with respect to the sparse set size.
To address this issue, it was noticed that for the particular structure of the squared exponential 2+3-body kernel, it is possible to map the mean prediction of a trained model onto an equivalent low-dimensional parametric model \cite{glielmo2018efficient, glielmo2020building, vandermause2020fly} without any loss of accuracy.
It was subsequently shown that the evaluation of the variance can also be mapped onto a low-dimensional model, achieving a dramatically accelerated uncertainty-aware BFF \cite{xie2021bayesian}.
While the 2-body and 3-body descriptions used in the previous work \cite{vandermause2020fly,xie2021bayesian} approach the computational speed of classical empirical potentials, they are limited in accuracy. To systematically increase the descriptive power of that approach, it is possible to extend the formalism to include higher body order interactions. However, inclusion of e.g. 4-body interactions requires summing contributions of all quadruplets of atoms in each neighborhood, which adds significant computational expense. Therefore, in a more recent work \cite{vandermause2021active} we used the atomic cluster expansion \cite{drautz_atomic_2019} to construct local structure descriptors that scale linearly with the number of neighbors, enabling efficient inclusion of higher body order correlations. An inner product kernel is then constructed based on the rotationally invariant ACE descriptors, forming a basis for a sparse Gaussian Process (GP) regression model for energies, forces and stresses. Crucially, it was shown in general that in the case of inner-product kernels with high dimensional many-body descriptors, the prediction of the mean can also be mapped exactly onto a constant-cost model via reorganization of the summation in the SGP mean calculation \cite{vandermause2021active}. This allows for efficient evaluation of the model forces, energies and stresses, but does not address the cost of evaluating uncertainties.
In this work, we present a method to map the variance of SGP models with inner product kernels. This advance enables large-scale uncertainty aware MD and overcome the linear scaling issue of SGPs.
Building on this approach, we achieve a significant acceleration of the Bayesian active learning (BAL) workflow and integrate it with Large-scale Atomic/Molecular Massively Parallel Simulator (LAMMPS)\cite{plimpton1995fast}.
Bayesian force fields are implemented within the LAMMPS MD engine, such that both forces and uncertainties for each atomic configuration are quantified at computational cost independent of the training set size. 

\ptitle{The objectives of the work. The justification for these objectives: Why is the work important?}
As a demonstration of the accelerated autonomous workflow, we train an uncertainty-aware many-body BFF for silicon carbide (SiC) on its several polymorphs and phases.
SiC is a wide-gap semiconductor with diverse applications ranging from efficient power electronics to nuclear physics and astronomy. 
With the discovery of a large number of extrasolar planets \cite{kim2022structure}, the compositions and processes under extreme conditions have led to a wide range of studies.
In particular, SiC has been identified from the adsorption spectroscopy of carbon-rich extrasolar planets \cite{madhusudhan2012possible,speck1997nature}, which has motivated numerous experimental and computational studies of its high temperature high pressure behavior.
The phase transition of SiC from the zinc blende (3C) to the rock salt (RS) phase is observed at high pressure in experiments \cite{tracy2019n,sekine1997shock,vogler2006hugoniot,miozzi2018equation,yoshida1993pressure,kidokoro2017phase,daviau2017zinc} and \textit{ab initio} calculations \cite{wang1996pressure,ran2021phase,lee2015first,durandurdu2004pressure,lu2008first,xiao2009ab,gorai2017pressure,kaur2020first}. 
Empirical potentials such as Tersoff\cite{tersoff_chemical_1994, erhart_analytical_2005}, Vashishta\cite{shimojo2000molecular, vashishta_interaction_2007}, MEAM\cite{kang_governing_2014}, and  Gao-Weber\cite{gao_empirical_2002} have been developed and applied in large-scale simulations for different purposes. 
However, empirical analytical potentials are limited in descriptive complexity, and hence accuracy, and require intensive human effort to select training configurations and to train. 
Machine learning approaches have allowed for highly over-parameterized or non-parametric models to be trained on a wide range of structures and phases \cite{handley2009optimal, bartok2018machine,zhang2018deep, batzner20223, musaelian2023learning, mailoa2019fast}.
Recently, neural network potentials were trained for SiC to study dielectric spectra\cite{chen2021study} and thermal transport properties\cite{fu2021deep}, but they do not capture high-pressure phase transitions.

In the present work, we deploy the accelerated autonomous BAL workflow to the high-pressure phase transition of SiC, and demonstrate that the transition process can be captured by the uncertainty quantification of the BFF.
Then the BFF is used to perform large-scale MD simulations, and compute vibrational and thermal transport properties of different phases. 
The FLARE BFF shows good agreement with ab-initio calculations\cite{ramakers2022effects} and experimental measurements, and significantly outperforms available empirical potentials in terms of accuracy, while retaining comparable computational efficiency.

\section{\label{sec:results}Results}


\subsection{Accelerated Bayesian Active Learning Workflow}


\ptitle{Active learning}
Directly using \textit{ab initio} MD to generate a sufficiently diverse training set for machine learning force fields is expensive and time consuming, and may still miss higher-energy configurations important for rare transformation phenomena. 
Here, we develop an active learning workflow, where MD is instead driven by the much faster surrogate FLARE many-body BFF. 
During the MD simulation, the model uses its internal uncertainty quantification, deciding to call DFT only when the model encounters atomic configurations with uncertainty above a chosen threshold. 
Within this framework, a much smaller number of DFT calls are needed, which greatly reduces the training time and increases the efficiency of phase space exploration. 

\ptitle{The main contribution}
In this work, we extend the formalism of efficient lossless mapping to include uncertainty of ACE-based many-body SGP models. Specifically, we implement mapped SGP variance to enable efficient uncertainty quantification and achieve large-scale MD simulations by interfacing the Bayesian active learning algorithm with LAMMPS.
The mappings of the forces and uncertainty overcome the scaling with the training set size of the computational cost of SGP regression, resulting in a significant acceleration of the training process in comparison with using the full SGP \cite{vandermause2021active}.

\begin{figure}[htbp]
    \centering
    \includegraphics[width=\textwidth]{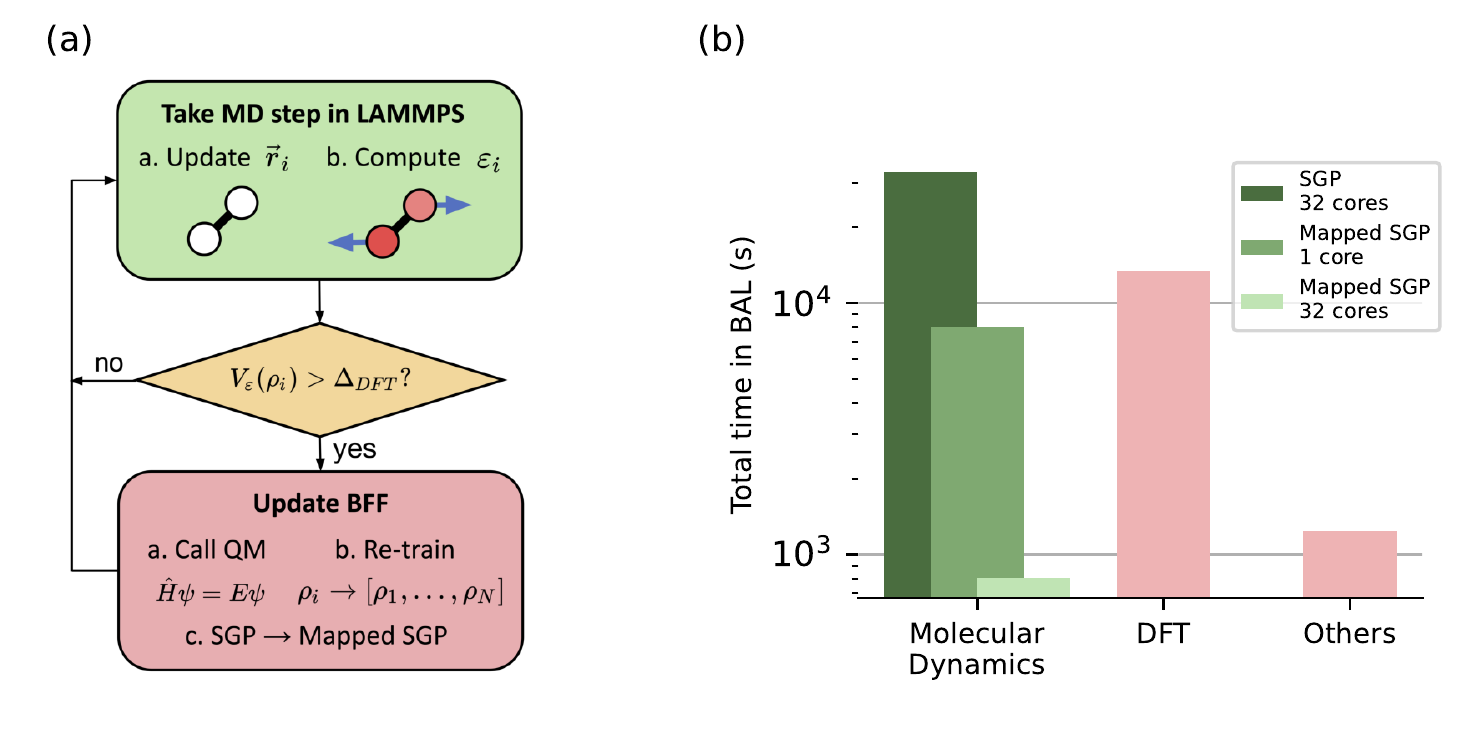}
    \caption{Workflow and performance. a. Bayesian active learning (BAL) workflow with LAMMPS. It closely follows our previous work\cite{vandermause2021active}, with the key addition that the SGP uncertainties are now mapped and therefore much cheaper, and the MD part of the training can be done within LAMMPS. b. Time profiling of the BAL workflow with a system of 72 atoms and 100,000 MD time steps. The SGP BAL workflow collected 14 frames with 72 atoms each, with 10 representative sparse environments selected per frame. The MD is greatly accelerated using mapped forces and uncertainty compared to sparse GP. ``Others" includes time consumed outside of MD and DFT, e.g. adding training data to SGP and optimizing hyperparameters.}
    \label{fig:wf}
\end{figure}

\ptitle{Describe each part of the workflow}
We illustrate our active learning workflow in Fig.~\ref{fig:wf}a. 
Starting from a SGP model with a small initial training set, we map both prediction mean and variance into quadratic models to obtain an efficient BFF (details are discussed in \textit{Methods}). 
With the mapped SGP force field, MD simulation runs in LAMMPS with uncertainty associated with the local energy assigned to each atom in a configuration at each time step. 
The MD simulation is interrupted once there are atoms whose uncertainties are above the threshold. 
Then DFT is used to compute energy, forces and stress for the high-uncertainty configurations. 
The training set is augmented with the newly acquired DFT data, the SGP model is retrained and mapped, and the MD simulation continues with the updated model.

\ptitle{Acceleration over previous implementation}
To illustrate the acceleration of active learning workflow with mapped forces and variance compared to that with the SGP model, we deploy a single active learning run of bulk 4H-SiC system and show the performance of each part of the workflow in Fig.~\ref{fig:wf}b. 
If the model is not mapped, the computational cost of MD with the SGP dominates the BAL procedure. 
When the mapped force field and variances are used, the computational cost is significantly reduced. 
The FLARE BFF achieves 0.76 ms$\cdot$CPU/step/atom in LAMMPS MD, which includes uncertainty quantification, and is comparable in speed to empirical interatomic potentials such as ReaxFF\cite{soria2018si}.
With mapped uncertainties, the DFT calculations become the dominant part from Fig.~\ref{fig:wf}b, i.e. the computational time for BAL is determined by the number of DFT calls.
We also note that the SGP model used for timing here has a small training set of 14 frames, and the computational cost of prediction grows linearly with sparse set size. 
It is worth emphasizing that after mapping, the prediction of model forces, energies, and stresses, and their variances is independent of the sparse set size. Therefore, the speedup versus the SGP model is more pronounced as more training data are collected and sparse representatives are selected.

\subsection{\label{sec:md}Bayesian Force Field for SiC High Pressure Phase Transition}


\ptitle{On-the-fly training settings, how does the uncertainty facilitate training (figure)}
In this work, we demonstrate the accelerated BAL procedure by training a mapped uncertainty-aware Bayesian force field to describe the phase transition of SiC at high pressure. We set up compressive and decompressive MD simulations at temperatures of 300 K and 2000 K for on-the-fly training simultaneously, where each SGP model is initialized with an empty training set.
The parameters set for the on-the-fly training workflow is in the Supplementary Table 1, and the complete training set information is in the Supplementary Table 2.

\begin{figure}[htbp]
    \centering
    \includegraphics[width=\textwidth]{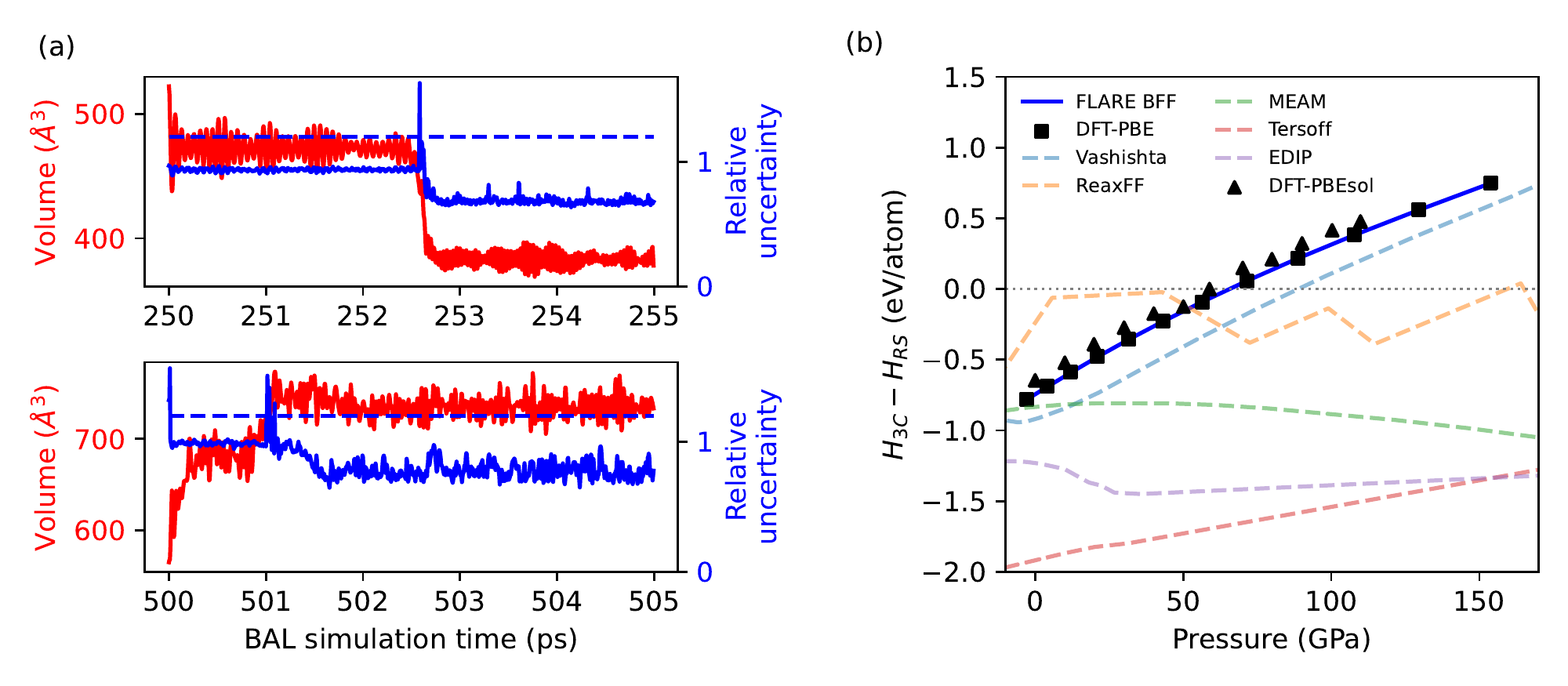}
    \caption{Phase transition simulation with FLARE BFF and enthalpy calculations. a. A 5 ps segment of the whole training trajectory where the 3C-RS phase transition is captured during the compressive (decompressive) on-the-fly active learning, the volume decreases (increases), and the model uncertainty spikes in the transition state. 
    The uncertainty threshold is shown as the blue dashed line. DFT is called, and new training data is added to the model when the relative uncertainty exceeds the threshold. 
    b. Enthalpy difference predictions from DFT (PBE, PBEsol\cite{lee2015first}), FLARE BFF and existing empirical potentials \cite{soria2018si,kang_governing_2014,tersoff_chemical_1994,lucas2009environment,shimojo2000molecular}
    at pressures from 0 to 150 GPa. The crossing with the dotted zero line gives the transition pressure predicted by enthalpy.}
    \label{fig:transition}
\end{figure}

The compressive MD starts with the 2H, 4H, 6H, and 3C polytypes that are stable at low pressure, and the pressure is increased by 30 GPa every 50 ps. 
The decompressive MD starts with the RS phase at 200 GPa, and the pressure is decreased by 20 GPa every 50 ps. The training data are collected by the BAL workflow shown in Fig.~\ref{fig:wf} . 
Fig.~\ref{fig:transition}a shows the system volume and relative uncertainty, i.e. the ratio between the uncertainty of the current frame and the average uncertainty of the training data set (see details in Supplementary Method 1\cite{supp}). 
In the compressive (decompressive) MD, when the transition happens, the volume decreases (increases) rapidly and the uncertainty spikes, since the model has never seen the transition state or the RS (3C) phase before. 
The post-transition high-pressure structure of the compression run has 6-fold (4-fold) coordination corresponding to the RS (3C) phase, and the transition is observed at 300 GPa (0 GPa) at room temperature. The difference of the transition pressures between the compressive and decompressive simulations is caused by nucleation-driven hysteresis. After the on-the-fly training is done, the training data from compressive and decompressive MD are combined to train a master force field for SiC, such that different phases at different pressures are covered by our force field.
In Supplementary Figure 2 and Supplementary Table 3\cite{supp}, we demonstrate the accuracy of our force field on cohesive energy and elastic constants in comparison with DFT.

\ptitle{enthalpy}
Using the mapped master force field, the phase transition pressure (at zero temperature) can be obtained from the enthalpy $H=E_\text{total} + PV$ of the different phases. At low pressure, the RS phase has higher enthalpy than 3C. With the pressure increased above 65 GPa, the enthalpy of RS phase becomes lower than 3C. 
As shown in Fig.\ref{fig:transition}, empirical potentials such as ReaxFF\cite{soria2018si}, MEAM\cite{kang_governing_2014}, Tersoff\cite{tersoff_chemical_1994} and EDIP\cite{lucas2009environment} produce qualitatively incorrect enthalpy curves, likely because they are trained only at low pressures. 
The Vashishta potential is the only one trained on the high pressure 3C-RS phase transition\cite{shimojo2000molecular} and presents a consistent scaling of enthalpy with pressures qualitatively, but it significantly overestimates the transition pressure at 90 GPa compared to DFT. 
FLARE BFF achieves a good agreement with the \textit{ab initio} (DFT-PBE) enthalpy predictions, with both methods yielding 65 GPa. We note that our PBE value is consistent with previous first-principles calculations such as 66.6(LDA)\cite{wang1996pressure} and 58(PBEsol)\cite{lee2015first}. 

\begin{figure}[htbp]
    \centering
    \includegraphics[width=\textwidth]{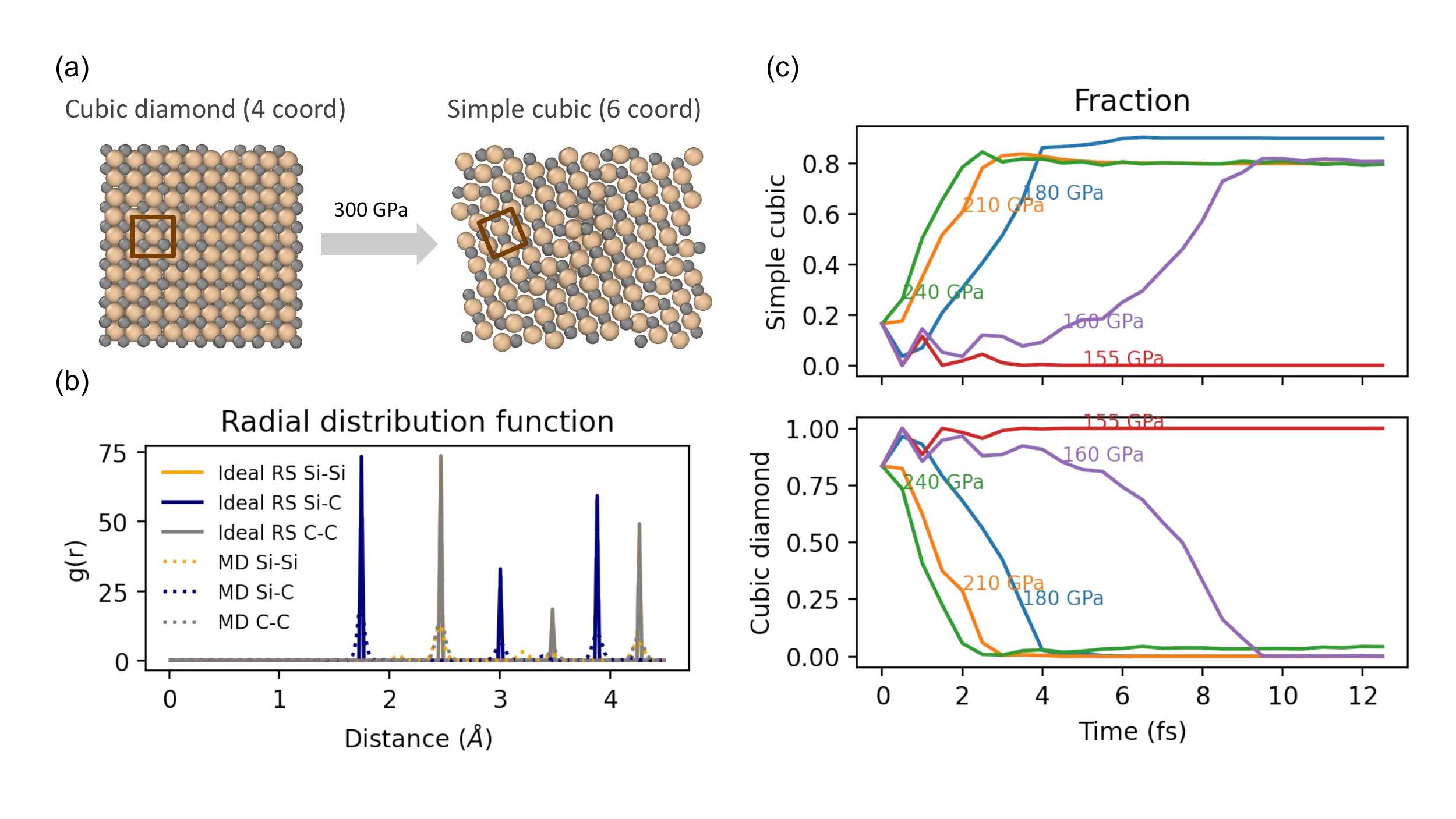}
    \caption{Large-scale simulation of phase transition and analysis. a. Large-scale MD with 1000 atoms for the 3C-RS phase transition after the training of FLARE BFF is finished. The transition is observed at 300 GPa at room temperature in MD.
    b. The radial distribution function of the final structure after transition in the MD, compared with the perfect RS crystal lattice, indicating the final configuration is in the RS phase. 
    c. The fraction of simple cubic (RS) and cubic diamond (3C, zinc-blende) atomic environments from the phase boundary evolution at different pressures.}
    \label{fig:phase-transition}
\end{figure}

\ptitle{large-scale MD result} 
Next we run a large-scale MD simulation with 1000 atoms for 500 ps at the temperature of 300 K. The NPT ensemble is used, and the pressure is increased by 30 GPa every 50 ps. At 300 GPa, the phase transition is observed from 4-fold coordination to 6-fold, as shown in Fig.\ref{fig:phase-transition}b. 
The final 6-fold coordinated structure has radial distribution function as shown in Fig.\ref{fig:phase-transition}c. The highest C-C, C-Si, Si-Si peaks match the perfect RS structure, confirming that the final structure is in the rock salt phase.

As in the smaller training simulations, the nucleation-controlled hysteresis caused the transition pressure (300 GPa) to be much higher than in experimental measurements (50-150 GPa) \cite{miozzi2018equation, tracy2019n,sekine1997shock}. 
To eliminate the nucleation barrier, we start our simulation with a configuration where the 3C and RS phases coexist, separated by a phase boundary, and evolve differently as a function of pressures. 
It is worth noting that such large scale two-phase simulations are enabled by the efficient force field and would not be possible to perform with DFT.
To recognize simple cubic (RS) and cubic diamond (3C, zinc-blende) environments, we use polyhedral template matching \cite{larsen2016robust} in OVITO \cite{stukowski2009visualization}.
The time evolution of the fractions of simple cubic and cubic diamond atomic environments is shown in Fig.~\ref{fig:phase-transition}c. The phase boundary evolves quickly within 10 fs. When pressure is greater than 160 GPa, the phase boundary evolves to RS phase, while below 155 GPa it evolves to 3C phase, which indicates the phase transition pressure is located between 155-160 GPa at the temperature of 300 K.

\begin{figure}[htbp]
    \centering
    \includegraphics[]{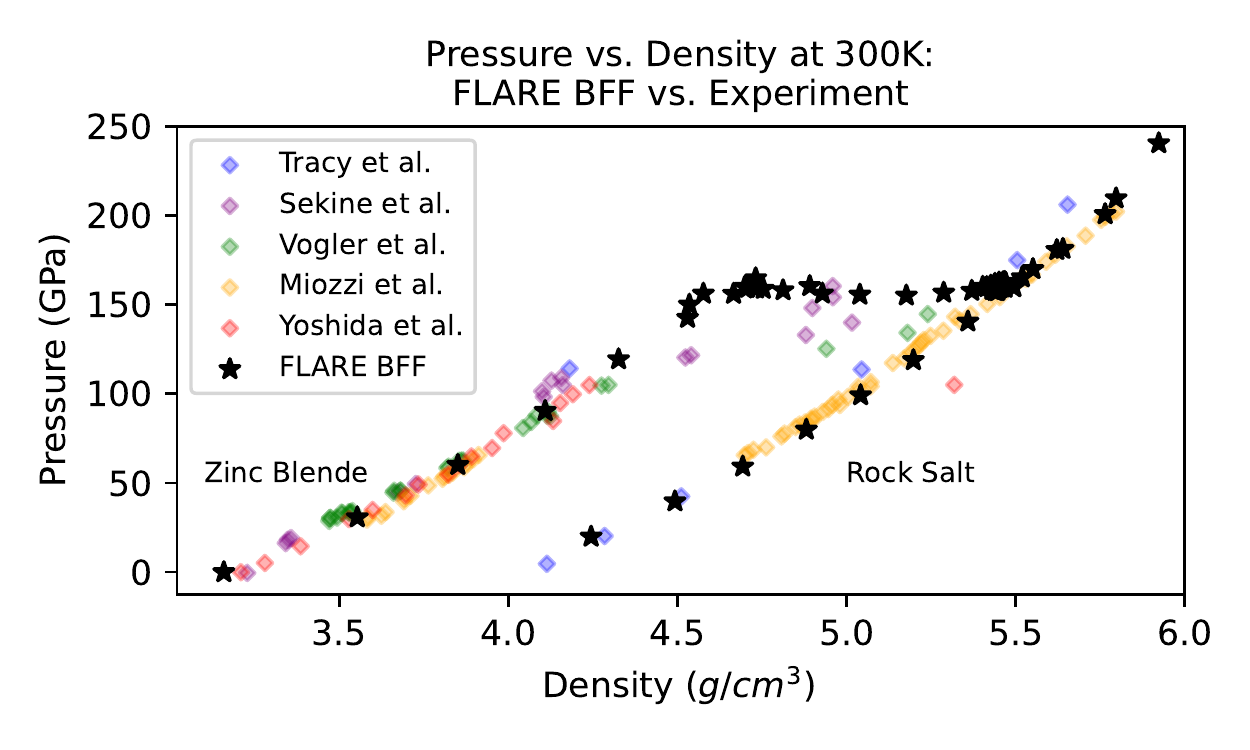}
    \caption{The density-pressure relation from experimental measurements\cite{tracy2019n,sekine1997shock,vogler2006hugoniot,miozzi2018equation,yoshida1993pressure} and MD simulations from FLARE BFF.  Zinc blende and rock salt phases correspond to two different lines of equation of state. Our FLARE BFF shows close agreement with experiments.}
    \label{fig:eos}
\end{figure}

A number of experiments have measured the density-pressure relations of zinc blende (3C) and rock salt at room temperature \cite{tracy2019n,sekine1997shock,vogler2006hugoniot,miozzi2018equation,yoshida1993pressure}.
As shown in Fig.~\ref{fig:eos}, the equation of state shows two parallel density-pressure curves, where the one with lower density is associated with the zinc-blende phase, and the other one with the rock salt phase. 
The pressures and corresponding densities at room temperature are extracted from the MD trajectories, and the equation of state is plotted to compare with the experimental measurements. 
FLARE BFF accurately agrees with the experimental measurements for both phases.
For the transition state, there are not many experimental data points available, and the measurements can be affected by the quality of the sample, but the FLARE BFF still shows good agreement with the measured data points.


\subsection{\label{sec:thermal}Vibrational and Thermal Transport Properties}

\begin{figure}[H]
    \centering
  \includegraphics[width=0.9\textwidth]{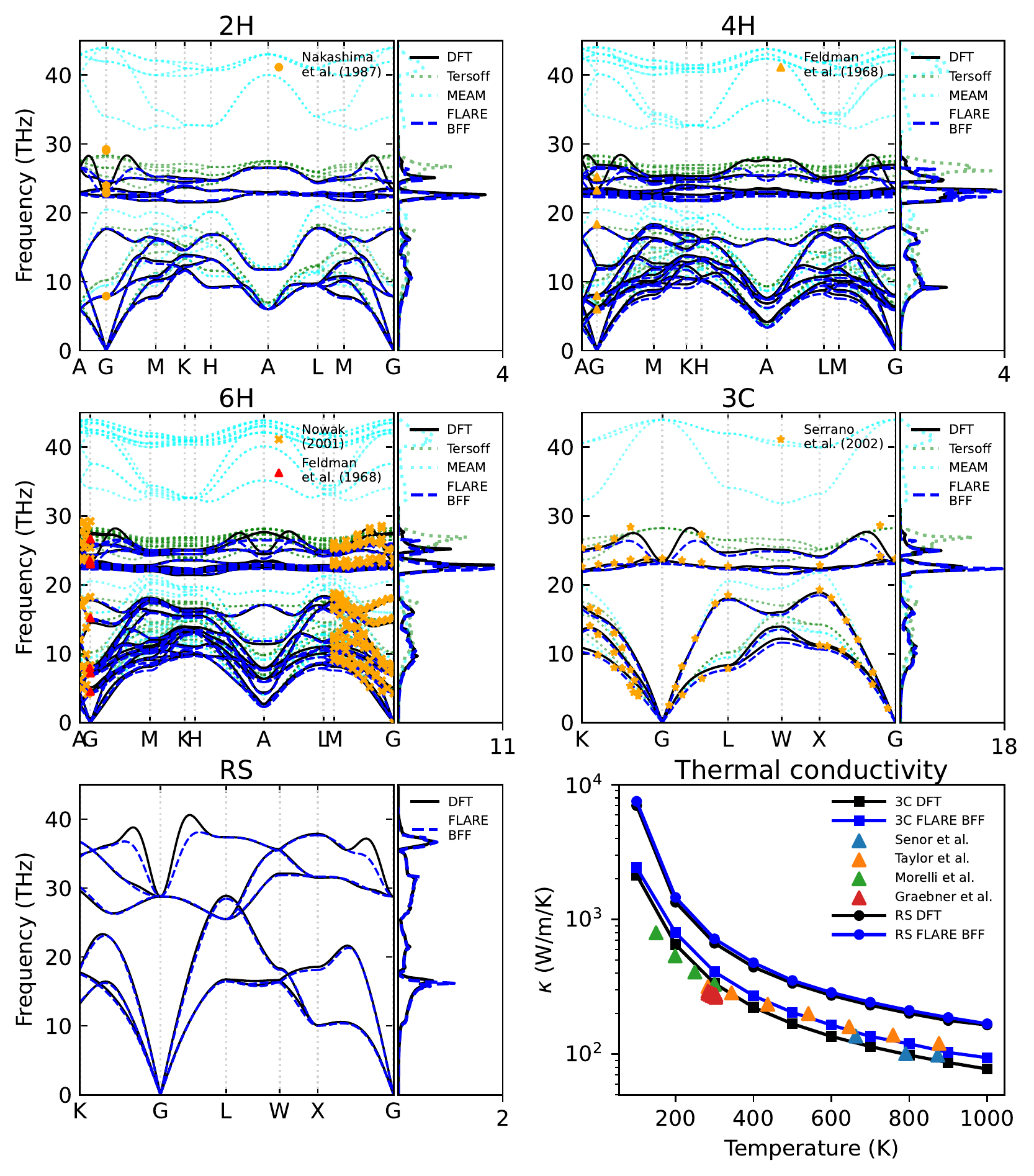}
    \caption{Phonon dispersions and phonon density of states of different polytypes from experimental measurements \cite{nowak2001crystal, feldman_phonon_1968, nakashima1987raman,serrano2002determination}, FLARE BFF and DFT calculations. Lower right: Thermal conductivity of 3C-SiC at 0 GPa and RS phase at 200 GPa with temperatures from 100 to 1000 K, from DFT and FLARE BFF calculations, and the experimental measurements\cite{taylor1993thermophysical,senor1996effects,morelli1994carrier,graebner1998report} are for 3C-SiC.}
    \label{fig:phonon}
\end{figure}

\ptitle{Thermal properties} 
To validate that the FLARE BFF gives accurate predictions of thermal properties, we investigate the phonon dispersions and thermal conductivities of the different polytypes and phases of SiC. The phonon dispersions are computed using Phonopy\cite{phonopy}. 
As shown in Fig.~\ref{fig:phonon}, FLARE BFF produces phonon dispersions in close agreement with both DFT and experiments\cite{nowak2001crystal, feldman_phonon_1968, nakashima1987raman,serrano2002determination} for both the low pressure polytypes and the high pressure rock salt phase at 200 GPa.
While in the optical branches the FLARE BFF prediction shows minor discrepancies with DFT at the highest frequencies, the phonon density of states (DOS) presents good agreement. 
In particular, FLARE BFF captures the peak at 23 THz corresponding to a number of degenerate optical branches. 
In Supplementary Figure 3, we show that the optical branches can be improved by increasing the cutoff of BFF.
Existing empirical potentials are much less accurate than FLARE BFF by comparing the phonon DOS in Fig.~\ref{fig:phonon}.
We note that the SiC crystals are polarized by atomic displacements and the generated macroscopic field induces an LO-TO splitting near $\Gamma$ point. 
The contribution from the polarization should be included through non-analytical correction (NAC) \cite{pick1970microscopic, phonopy}, and the first principle calculation with NAC is discussed in Ref.\cite{protik_phonon_2017}. However, since FLARE BFF model does not contain charges and polarization, the NAC term is not considered. 
Thus, our comparison is made between the FLARE BFF and DFT without NAC. 
The lack of NAC accounts for the disagreement in the high frequencies of the optical bands of 2H DFT phonon (Fig.~\ref{fig:phonon}) with experimental measurements near $\Gamma$ point.

\ptitle{Thermal conductivity.} 
Having confirmed the accuracy of the 2nd order force constants by the phonon dispersion calculations, we then compute the thermal conductivity within the Boltzmann transport equation (BTE) formalism. The 2nd and 3rd order force constants are computed using the Phono3py\cite{phono3py} code, and then used in the Phoebe\cite{cepellotti2022phoebe} transport code to evaluate thermal conductivity with the iterative BTE solver\cite{omini1995iterative}. 
In Supplementary Figure 4\cite{supp}, we verify that the exclusion of the non-analytic correction does not have a significant effect on the thermal conductivity values.
Fig.~\ref{fig:phonon} presents the thermal conductivities of the zinc blende phase at 0 GPa and the rock salt phase at 200 GPa as a function of temperature.
The FLARE BFF results are in good agreement with the DFT-derived thermal conductivity for both zinc blende and rock salt phases.
The thermal conductivity of zinc blende phase computed from DFT and FLARE BFF is also in good agreement with experimental measurements\cite{taylor1993thermophysical,senor1996effects,morelli1994carrier,graebner1998report}.
For the high-pressure rock salt phase, the thermal conductivity has not been previously computed or measured to the best of our knowledge. 
Therefore, our calculation provides a prediction that awaits experimental verification in the future.

\section{\label{sec:discussions}Discussion}


\ptitle{In conclusion, developed a method}
In this work we develop a BFF that maps both the mean predictions and uncertainties of SGP models for many-body interatomic force fields. 
The mapping procedure overcomes the linear scaling issue of SGPs and results in near-quantum accuracy, while retaining computational cost comparable to empirical interatomic potentials such as ReaxFF.
The efficient uncertainty-aware BFF model forms the basis for the construction of an accelerated autonomous BAL workflow that is coupled with the LAMMPS MD engine and enables large-scale parallel MD simulations. The key improvement with respect to previous methods is the ability of the many-body model to efficiently calculate forces and uncertainties at comparable computational cost.

\ptitle{Application}
As a demonstration of the ability of this method to capture and learn subtle interactions driving phase transformations on-the-fly, we use the BAL workflow to train a BFF for SiC on several common polytypes and phases. 
The zinc-blende to rock-salt transition is captured in both the active learning and the large-scale simulation, facilitated by the model uncertainty.
FLARE BFF is shown to have good agreement with DFT for the enthalpy prediction in a wide range of pressure values. The BFF model is readily employed to perform a large-scale MD of the phase boundary evolution, which allows for reliable identification of the transition pressure at room temperature to be located at 155-160 GPa. The density-pressure relation predicted by FLARE BFF agrees very well with experimental measurements. 
We also find close agreement for phonon dispersions and thermal conductivities of several SiC phases, as compared with DFT calculations and experiments, outperforming existing empirical potentials.


The high-performance implementation of BFFs, combining accuracy with autonomous uncertainty-driven active learning, opens numerous possibilities to explicitly study dynamics and microscopic mechanisms of phase transformations and non-equilibrium properties such as thermal and ionic transport. The presented unified approach can be extended to a wide range of complex systems and phenomena, where interatomic interactions are difficult to capture with classical approaches while time- and length-scales are out of reach of first-principles computational methods.
Uncertainty quantification in MD simulations allows for systematic monitoring of the model confidence and detection of rare and unanticipated phenomena, such as reactions or nucleation of phases. Such events are statistically unlikely to occur in smaller simulations and become increasingly likely and relevant as the simulation sizes increase, as may be needed to study complex and heterogeneous materials systems.

\section{\label{sec:methods}Methods}
In this section, we use bold letters such as $\boldsymbol{\alpha}$ and $\mathbf{\Lambda}$ to denote a vector or a matrix, and use indices such as $\alpha_i$ and $\Lambda_{ij}$ to denote the components of the vector and matrix respectively.
In addition, we use $F$ to represent all the collected configurations with all atomic environments (the ``full'' data set), and use $S$ for a subset atomic environments selected from those configurations (the ``sparse'' data set). 

\subsection{Gaussian Process Regression}

\ptitle{What is atomic environment. What is GP. what is kernel}
The local atomic environment $\rho_i$ of an atom $i$ consists of all the neighbor atoms within a cutoff radius, and is associated with a label $y_i$ that can be the force $\mathbf{F}_i$ on atom $i$ or a local energy $\varepsilon_i$. 
The local energy labels are usually not available in practice, but total energies, stresses and atomic forces are.
A kernel function $k(\rho_i, \rho_j)$ quantifies the similarity between two atomic environments, which is also the covariance between local energies. 
Gaussian process regression (GP) assumes a Gaussian joint distribution of all the training data $\{(\rho_i, y_i)\}$ and test data $(\rho, y)$.
The posterior distribution for test data is also Gaussian with the mean and variance 
\begin{align}
    \varepsilon(\rho) &= \mathbf{k}_{\varepsilon F} (\mathbf{K}_{FF}+\mathbf{\Lambda})^{-1} \mathbf{y}\\
    V(\rho) &= k_{\varepsilon \varepsilon} - \mathbf{k}_{\varepsilon F} (\mathbf{K}_{FF}+\mathbf{\Lambda})^{-1}\mathbf{k}_{F \varepsilon}
\end{align}
Here, $\varepsilon$ is the local energy of $\rho$, $\mathbf{k}_{\varepsilon F}$ is the kernel vector describing covariances between the test data and all the training data, with element $k(\rho, \rho_i)$, $k_{\varepsilon \varepsilon}=k(\rho, \rho)$ is the kernel between test data $\rho$ and itself, $\mathbf{K}_{FF}$ is the kernel matrix with element $k(\rho_i, \rho_j)$, $\mathbf{\Lambda}$ is a diagonal matrix describing noise, and $\mathbf{y}$ is the vector of all training labels. 

\ptitle{What is SGP: what assumptions are made, what are the posterior mean and var}
Since the full GP evaluates the kernel between all configurations in the whole training set, it becomes computationally inefficient for large data sets. A sparse approximation is needed such that the computational cost can be reduced while keeping the information as complete as possible. 
Following our recent work \cite{vandermause2021active} we consider the mean prediction from Deterministic Training Conditional (DTC) Approximation \cite{quinonero2005unifying}, and the variance on local energies
\begin{align}
    \varepsilon(\rho) &= \mathbf{k}_{\varepsilon S} (\mathbf{K}_{SF}\mathbf{\Lambda}^{-1}\mathbf{K}_{FS} + \mathbf{K}_{SS})^{-1} \mathbf{K}_{SF} \mathbf{y} \label{eq:sgp_mean}\\
    V(\rho) &= k_{\varepsilon\varepsilon} - \mathbf{k}_{\varepsilon S} K_{SS}^{-1} \mathbf{k}_{S\varepsilon} \label{eq:sgp_var}
\end{align}
where the $\mathbf{k}_{\varepsilon S}$ is the kernel vector between the test data $\rho$ and the sparse training set $S$, $\mathbf{K}_{SF}$ is the kernel matrix between the sparse subset $S$ and the complete training set $F$, and $\mathbf{K}_{SS}$ is the kernel matrix between $S$ and itself.
The energy, forces and stress tensor of a test configuration are given by the mean prediction, and their corresponding uncertainties are given by the square root of the predictive variance.

\subsection{ACE Descriptors with Inner Product Kernel}
\ptitle{intro of ace} In the FLARE SGP formalism \cite{vandermause2021active} we use the atomic cluster expansion\cite{drautz_atomic_2019} (ACE) descriptors to represent the features of a local environment. The total energy is constructed from atomic clusters, represented by the expansion coefficients of an atomic density function using spherical harmonics. 
We refer the readers to Ref.\cite{drautz_atomic_2019,vandermause2021active} for more details.
To build up the SGP model for interatomic Bayesian force field, or BFF, we use the inner product kernel defined as
\begin{equation}
    k(\rho_1, \rho_2) = \sigma^2 \left(\frac{\mathbf{d}_1 \cdot \mathbf{d}_2}{d_1 d_2} \right)^\xi
\end{equation}
where $\mathbf{d}_1$ and $\mathbf{d}_2$ are ACE descriptors of environments $\rho_1$ and $\rho_2$, $\sigma$ is the signal variance which is optimized by maximizing the log likelihood of SGP, and $\xi$ is the power of the inner product kernel which is selected a priori. 
We also normalize the kernel by the L2 norm of the descriptors. For simplicity of notations and without loss of generality, we ignore the normalization and only showcase the kernel without derivative associated with energy.

\ptitle{Force Mapping and Variance Mapping}
With inner product kernels, a highly efficient but lossless approximation is available via reorganization of the summation in the mathematical expression\cite{vandermause2021active}.
Defining $\boldsymbol{\alpha}:=(\mathbf{K}_{SF}\mathbf{\Lambda}^{-1}\mathbf{K}_{FS} + \mathbf{K}_{SS})^{-1} \mathbf{K}_{SF} \mathbf{y}$ and $\tilde{\boldsymbol{d}_i}=\mathbf{d}_i/d_i$, the mean prediction of test data $\rho_i$ (Eq.\ref{eq:sgp_mean}) from sparse training set $S=\{\rho_s\}$ can be written as \cite{vandermause2021active}
\begin{equation}
\begin{split}
\varepsilon(\rho_i) &= \sigma^2 \sum_t (\tilde{\boldsymbol{d}_i} \cdot \tilde{\boldsymbol{d}_t})^\xi \alpha_t \\
&= \sigma^2 \sum_{t, m_1, ..., m_{\xi}} \tilde{d}_{im_1} \tilde{d}_{tm_1} \cdot\cdot\cdot \tilde{d}_{i m_\xi} \tilde{d}_{tm_\xi} \alpha_t \\
&= \sigma^2 \sum_{m_1, ..., m_\xi} \tilde{d}_{im_1} \cdot \cdot \cdot \tilde{d}_{im_\xi} \left( \sum_t \tilde{d}_{t m_1} \cdot \cdot \cdot \tilde{d}_{t m_\xi} \alpha_t \right) \\
&= \sum_{m_1, ..., m_\xi} \tilde{d}_{im_1} \cdot\cdot\cdot \tilde{d}_{im_\xi} \beta_{m_1, ..., m_\xi},
\end{split}
\label{eq:force_map}
\end{equation}
The $\boldsymbol{\beta}$ tensor can be computed from the training data descriptors. Thus, we can store the $\boldsymbol{\beta}$ tensor, and during the prediction we directly evaluate Eq.~\ref{eq:force_map} without the need to sum over all training data $t$. 

\ptitle{Present variance mapping derivation}
Crucially, the variance Eq.~\ref{eq:sgp_var} has the similar reorganization as the mean prediction of Eq.~\ref{eq:force_map}

\begin{align}
V(\rho_i) &= \sigma^2 (\tilde{\boldsymbol{d}}_i \cdot \tilde{\boldsymbol{d}}_i)^\xi 
- \sigma^4 \sum_{s,t} (\tilde{\boldsymbol{d}}_i \cdot \tilde{\boldsymbol{d}}_s)^\xi (K_{SS}^{-1})_{st} (\tilde{\boldsymbol{d}}_t \cdot \tilde{\boldsymbol{d}}_i)^\xi\\
&= \sum_{\substack{m_1, ..., m_\xi \\ n_1,...,n_\xi}} (\tilde{d}_{im_1} \cdot\cdot\cdot \tilde{d}_{im_\xi}) \gamma_{\substack{m_1, ..., m_\xi \\ n_1,...,n_\xi}} (\tilde{d}_{in_1} \cdot\cdot\cdot \tilde{d}_{in_\xi})
\label{eq:var_map}
\end{align}
where $\boldsymbol{\gamma}$ is a tensor that can be calculated and stored once the training data is collected, and used in inference without explicit summation over training data $i$.

The reorganization indicates that the SGP regression with inner product kernel essentially gives a polynomial model of the descriptors. Denote $n_d$ as the descriptor dimension. The reorganized mean prediction has both the $\boldsymbol{\beta}$ size and the computational cost as $O(n_d^\xi)$.
While the reorganized variance prediction has both the $\boldsymbol{\gamma}$ size and the computational cost as $O(n_d^{2\xi})$.
By the reorganization, both the mean and variance predictions become independent of the training set and get rid of the linear scaling with respect to training size.
Here, we choose $\xi=2$ for the mean prediction, because 1) it is shown \cite{vandermause2021active} that $\xi=2$ has a significant improvement of likelihood compared to $\xi=1$, while the improvement of $\xi>2$ is marginal; and 2) higher order requires a much larger memory for $\boldsymbol{\beta}$ tensor, and the evaluation of Eq.\ref{eq:force_map} is much costlier than $\xi=2$.

From Eq.\ref{eq:var_map} we see that the variance has twice the polynomial degree of that for the mean prediction. For example, when $\xi=2$, the mean prediction is a quadratic model while the variance is a quartic model of descriptors. For computational efficiency, in this work we use $V_{\xi=1}$ for variance prediction. $V_{\xi=1}$ is not the exact variance of our mean prediction $\varepsilon_{\xi=2}$, but an approximation using the same hyperparameters as $V_{\xi=2}$. It has a strong correlation with the exact variance $V_{\xi=2}$ as shown in Supplementary Figure 1\cite{supp}.
Especially, the correlation coefficients between $V_{\xi=1}$ and higher powers are close to 1.0, the perfect linear relation, even with a training size of 200 frames. 
This indicates that uncertainty quantification with $V_{\xi=1}$ is able to recognize discrepancies between different configurations as well as $V_{\xi=2}$. Specifically, atomic environments with higher $V_{\xi=2}$ uncertainties will also be assigned higher $V_{\xi=1}$ uncertainties than others. It is then justified that $V_{\xi=1}$ can be used as a strongly correlated but much cheaper approximation of $V_{\xi=2}$.

\ptitle{summarize} To summarize, we use $\varepsilon_{\xi=2}$ and $V_{\xi=1}$ for mean and variance predictions respectively, where both are quadratic models with respect to the descriptors.
During the on-the-fly active learning, every time the training set is updated by the new DFT data, the $\boldsymbol{\beta}$ and $\boldsymbol{\gamma}$ matrices are computed from training data descriptors and stored as coefficient files. In LAMMPS MD, the quadratic models are evaluated to make predictions for energy, forces, stress and uncertainty of the configurations.  

\section{Data Availability}
The scripts and data are available in Zenodo: \url{https://doi.org/10.5281/zenodo.5797177}.

\section{Code Availability}

The code for FLARE Bayesian force field and Bayesian active learning is available at \url{https://github.com/mir-group/flare}.

\section{Acknowledgement}
We thank Lixin Sun, Thomas Eckl, Matous Mrovec, Thomas Hammerschmidt and Ralf Drautz for helpful discussions on the construction of the force field and properties of SiC. We thank Jenny Coulter and Andrea Cepellotti for the helpful instructions on thermal conductivity calculation using Phono3py and Phoebe. We thank Jenny Hoffman, Sushmita Bhattacharya and Kristina Jacobsson for constructive suggestions on the manuscript. 
YX was supported from the US Department of Energy (DOE), Office of Science, Office of Basic Energy Sciences (BES) under Award No.\ DE-SC0020128. JV was supported by the National Science Foundation award number 2003725. AJ was supported by the Aker scholarship. BK was supported by the National Science Foundation through the Harvard University Materials Research Science and Engineering Center under award number DMR-2011754 and Robert Bosch LLC.
Computational resources were provided by the FAS Division of Science Research Computing Group at Harvard University.
\section{Author contributions}
J.V., Y.X. and A.J. developed the FLARE code, the corresponding LAMMPS pair styles and compute commands. Y.X. implemented the on-the-fly training workflow coupled with LAMMPS. 
Y.X. performed the on-the-fly training and MD simulation of SiC. S.R. performed DFT calculations of SiC polymorphs.
Y.X. computed the thermal conductivity of SiC with the contribution of N.H.P.. 
B.K. conceived the application and supervised the work. 
Y.X. and B.K. wrote the manuscript. All authors contributed to manuscript preparation.

\section{Competing interests}
The authors declare no competing financial or non-financial interests.

\bibliographystyle{unsrt}
\bibliography{main}

\end{document}


\title{Uncertainty-aware Molecular Dynamics from Bayesian Active Learning for Phase Transformations and Thermal Transport in SiC}

\author{Yu Xie}
\email{xiey@g.harvard.edu}
\affiliation{John A. Paulson School of Engineering and Applied Sciences, Harvard University}
 
\author{Jonathan Vandermause}
\affiliation{John A. Paulson School of Engineering and Applied
Sciences, Harvard University}
\affiliation{Department of Physics, Harvard University}

\author{Senja Ramakers}
\affiliation{Corporate Sector Research and Advance Engineering, Robert Bosch GmbH, Renningen}
\affiliation{Interdisciplinary Centre for Advanced Materials Simulation, Ruhr-Universit\"at Bochum}

\author{Nakib H. Protik}
\affiliation{John A. Paulson School of Engineering and Applied Sciences, Harvard University}

\author{Anders Johansson}
\affiliation{John A. Paulson School of Engineering and Applied Sciences, Harvard University}

\author{Boris Kozinsky}
\email{bkoz@seas.harvard.edu}
\affiliation{John A. Paulson School of Engineering and Applied Sciences, Harvard University}
\affiliation{Robert Bosch LLC, Research and Technology Center}


\maketitle
\begin{center}
\vspace{-30pt}
    \Large Supplementary Materials
\end{center}

\toccontents

\newpage
\section*{Supplementary Methods}
\subsection*{Supplementary Method 1: Select Uncertainty Threshold by Training Set Uncertainty\label{supp-method}}

In the active learning, the Bayesian force field predicts uncertainty to quantify the confidence of the model of the current frame. Then the uncertainty is compared to an  acquisition threshold. If the uncertainty is higher than the threshold then a DFT call is made to compute the ``ground truth'' energy (e), forces (f) and stress (s) of the current configuration. 
The uncertainties of total potential energy, forces and stress are computed from the square root of the variance of the full GP:
\begin{align}
   V_{pp}(\rho) &= k_{pp} - \mathbf{k}_{p F}^\top (\mathbf{K}_{FF}+\mathbf{\Lambda})^{-1}\mathbf{k}_{Fp},\quad p=e, f, s
   \label{supp-eq:gp_var}
\end{align}
which depend on the signal variance of the kernel function and the noise parameters $\sigma_n$. Therefore, choosing an absolute (constant) threshold throughout the active learning ignores the dependence and may determine a frame as ``familiar/unfamiliar" inaccurately.
In Ref.\cite{vandermause2020fly,xie2021bayesian}, a multiple of $\sigma_n$is used as the threshold, such that the threshold changes while the noise parameter is optimized to a different value. In Ref.\cite{vandermause2021active}, a simple form of the raw uncertainty of sparse GP is used 
\begin{align}
    V(\rho) &= \frac{1}{\sigma^2}(k_{\varepsilon\varepsilon} - \mathbf{k}_{\varepsilon S} K_{SS}^{-1} \mathbf{k}_{S\varepsilon}) 
    \label{supp-eq:sgp_var}
\end{align}
where the normalization by the signal variance $\sigma$ make $V(\rho)$ independent of the signal variance and noise. Then an absolute (constant) threshold can be used in this special case.

In this work, we introduce a more general way to determine the threshold. Particularly, we compute the average uncertainty $\bar{V}_\text{train}$ of all the training data, and then use a multiple of $\bar{V}_\text{train}$ as a threshold. In the active learning of SiC, we call the DFT when $\sqrt{\bar{V}} > 1.5 \sqrt{\bar{V}_\text{train}}$.
The relative uncertainty plotted in the Fig.2a in main text is $\sqrt{\bar{V}/\bar{V}_\text{train}}$.

\newpage
\section*{Supplementary Tables}
\subsection*{Supplementary Table \ref{supp-tab:otf}: On-the-fly Training Settings}
\begin{table}[H]
\begin{tabular}{ccc}
Parameter	& Value		&	Description \\
\hline
cutoff ($\text{\AA}$)           & 5.0  & Cutoff radius of the local environment for ACE descriptors          \\
$\sigma^\text{(B1)}$              & 78.2991                    & Signal variance of kernel of B1 descriptors                         \\
$\sigma^\text{(B2)}$              & 3.5271                     & Signal variance of kernel of B2 descriptors                         \\
$\sigma_\text{n}^\text{(energy)}$ & 0.1086                     & Noise parameter for total energy                                    \\
$\sigma_\text{n}^\text{(forces)}$ & 0.0895                    & Noise parameter for forces                                          \\
$\sigma_\text{n}^\text{(stress)}$ &  0.0021                  & Noise parameter for stress                                          \\
$n_\text{max}^\text{(B1)}$        & 11 & Maximal index of radial basis function of ACE B1 descriptors        \\
$n_\text{max}^\text{(B2)}$        & 8  & Maximal index of radial basis function of ACE B2 descriptors        \\
$l_\text{max}^\text{(B2)}$        & 4  & Maximal index of spherical harmonics function of ACE B2 descriptors\\
\hline
\end{tabular}
    \caption{For the details of the parameters of the kernels, we direct the reader to Ref.\cite{vandermause2021active}. For the details of the parameters of the ACE descriptors, we direct the reader to Ref.\cite{drautz_atomic_2019}}
    \label{supp-tab:otf}
\end{table}

\newpage
\subsection*{Supplementary Table \ref{supp-tab:training_set}: Training Set Collected from Bayesian Active Learning}

\begin{table}[H]
\begin{tabular}{ccccccccccc}
 \hline
 System &	Pressure  &	Temperature  &	$\tau_{\text{sim}}$  &	$\tau_{\text{wall}}$ &	$N_\text{nodes}$ &	$N_{\text{atoms}}$ &	$N_{\text{struc}}$ &	$N_{\text{envs}}$ &	$N_{\text{sparse}}$ &	$N_{\text{labels}}$\\
 \hline
3C    & 0-480 & 2000                  & 0.85                  & 20.94  & 1                     & 64                    & 110 & 7040  & 550  & 21890  \\
2H    & 0-480 & 2000                  & 0.85                  & 34.94  & 1                     & 72                    & 125 & 9000  & 625  & 27875  \\
4H    & 0-480 & 2000                  & 0.85                  & 36.87  & 1                     & 72                    & 131 & 9432  & 655  & 29213  \\
6H    & 0-480 & 2000                  & 0.85                  & 74.04  & 1                     & 108                   & 122 & 13176 & 610  & 40382  \\
3C    & 0-800 & 300                   & 0.85                  & 9.78   & 1                     & 64                    & 72  & 4608  & 360  & 14328  \\
RS    & 200-0 & 2000                  & 0.55                  & 17.22  & 1                     & 64                    & 120 & 7680  & 600  & 23880  \\
Total & -     & - & - & 193.79 & - & - & 680 & 50936 & 3400 & 157568\\
 \hline
\end{tabular}
\caption{Summary of on-the-fly training for the SiC system: the simulation temperature $T$ (K); the pressure (GPa) range of the simulation; the total simulation time $\tau_{\text{sim}}$ (ns); the total wall time of the simulation $\tau_{\text{wall}}$ (hr); the number of CPU nodes used for the training (with 32 CPUs per node); the number of atoms in the simulation $N_{\text{atoms}}$; the total number of structures added to the training set $N_{\text{struc}}$; the total number of local environments in all training structures $N_\text{envs}$; the total number of local environments added to the sparse set $N_{\text{sparse}}$; and the total number of training labels $N_{\text{labels}}$.}
\label{supp-tab:training_set}
\end{table}

We note that since each on-the-fly training only uses a single CPU node with 32 cores, in practice we run these on-the-fly training simulations simultaneously on a cluster. 
The ``System" in the table specifies the initial polytype/phase of the simulation. The simulations starting with 3C/2H/4H/6H are compressive with increasing pressures. The simulation starting with RS is decompressive with decreasing pressures. 
The pressure range in the table also specifies the compression/decompression. For example, ``0-480'' means a compressive simulation from 0 to 480 GPa, and ``200-0'' means a decompressive simulation from 200 to 0 GPa.

\newpage
\subsection*{Supplementary Table \ref{supp-tab:elastic}: Elasticity}

\begin{table}[htbp]
\renewcommand{\arraystretch}{0.9}
\begin{threeparttable}
\footnotesize{
\begin{tabular}{cccccccccc}
\hline
Polytype &  Method   & $C_{11}$ & $C_{33}$  & $C_{12}$  & $C_{13}$  & $C_{44}$  & $C_{66}$  & $B_{\text{V}}$  & $G_{\text{V}}$  \\
\hline
\multirow{5}{*}{3C} & EXP & 390\tnote{a}, 395\tnote{e}, 371\tnote{c} & & 142\tnote{a}, 123\tnote{e}, 146\tnote{c} &                & 150\tnote{b}, 256\tnote{a}, 236\tnote{e},111\tnote{c} &                &                      &                      \\
                    & DFT & 384.3                                    & 384.3          & 127.9                                    & 127.9          & 239.9                                                 & 239.9          & 213.6                & 195.4                \\
                    &  FLARE BFF  &  373.1  &  373.1  &  128.0  &  128.0  &  230.3  &  230.3  &  209.7  &  187.2\\
                    & Tersoff & 426   & 426   & 109.9 & 109.9 & 252.7 & 252.7 & 215.3 & 214.8 \\
                    & MEAM    & 402.2 & 402.2 & 115.6 & 115.6 & 215.1 & 215.1 & 211.1 & 186.4 \\
\hline
\multirow{5}{*}{2H} & EXP &                                          &                &                                          &                &                                                       &                &                      &                      \\
                    & DFT & 494.3                                    & 533.5          & 102.1                                    & 50.7           & 151.3                                                 & 196.1          & 214.3                & 187.6                \\
                    &  FLARE BFF  &  482.8  &  527.2  &  101.0  &  52.7  &  158.1  &  190.9  &  211.7  &  187.2\\
                    & Tersoff & 502.2 & 541.6 & 84.4  & 42.3  & 188.8 & 208.9 & 209.3 & 215.9 \\
                    & MEAM    & 467.4 & 497.9 & 98.2  & 67.7  & 167.2 & 184.6 & 211.1 & 189.1 \\
                    \hline
\multirow{5}{*}{4H} & EXP & 501\tnote{d}                             & 553\tnote{d}   & 111\tnote{d}                             & 52\tnote{d}    & 163\tnote{d}                                          &                &                      &                      \\
                    & DFT & 487.8                                    & 533.4          & 105                                      & 51.9           & 157.9                                                 & 191.4          & 214                  & 188.1                \\
                    &  FLARE BFF  &  476.0  &  514.9  &  104.8  &  55.0  &  158.4  &  185.6  &  210.7  &  184.0\\
                    & Tersoff & 506.6 & 547.7 & 86.6  & 44.9  & 189.1 & 210   & 212.7 & 216.9 \\
                    & MEAM    & 467.2 & 497.9 & 98.4  & 67.7  & 167.2 & 184.4 & 211.1 & 189.1 \\
                    \hline
\multirow{5}{*}{6H} & EXP & 501\tnote{d}                             & 553\tnote{d}   & 112\tnote{d}                             & 52\tnote{d}    & 163\tnote{d}                                          &                &                      &                      \\
                    & DFT & 485.2                                    & 534.1          & 106                                      & 52.3           & 160.1                                                 & 189.6          & 213                  & 188.2                \\
                    &  FLARE BFF  &  472.2  &  515.3  &  106.1  &  55.3  &  158.4  &  183.1  &  210.3  &  182.8\\
                    & Tersoff & 507.2 & 548.6 & 87    & 45.3  & 189.1 & 210.1 & 213.1 & 217   \\
                    & MEAM    & 467.2 & 497.9 & 98.4  & 67.7  & 167.2 & 184.4 & 211.1 & 189   \\
                    \hline
\end{tabular}
}
\begin{tablenotes}[para, online]\footnotesize
\item[a] Ref. \cite{feldman_phonon_1968}
\item[b] Ref. \cite{harrison_electronic_1981}
\item[c] Ref. \cite{pestka_measurement_2008}
\item[d] Ref. \cite{kamitani_elastic_1997}
\item[e] Ref. \cite{djemia_elastic_2004}
\item[] (Unit: GPa)
\end{tablenotes}
\end{threeparttable}
\caption{Elastic tensor ($C_{11}\sim C_{66}$), bulk modulus ($B_V$) and shear modulus ($G_V$) from experiments (EXP), DFT\cite{ramakers2022effects}, FLARE BFF and empirical potentials (Tersoff, MEAM).}
\label{supp-tab:elastic}
\end{table}

We note that since FLARE BFF is trained on DFT data but not the experiments, we expect it to give predictions closer to DFT than experiments, while as shown in Table S\ref{supp-tab:elastic}, we still find good match for the four polytypes between FLARE BFF and experimental results. 

\newpage
\subsection*{Supplementary Table \ref{sup-tab:speed_compare}: Performance Comparison}

\begin{center}
\begin{table}[htbp]
    \newcolumntype{P}[1]{>{\centering\arraybackslash}p{#1}}
    \centering
    \begin{tabular}{P{2.3cm}P{5.5cm}P{2.1cm}P{3.1cm}P{1.5cm}}
    \hline
      Model   &  Parameters & Speed\ \ \ \ \ \  (potential) & Speed\ \ \ \ \ \ \ \ \ \ \ \ (variance) & Force MAE \\
      \hline
      2+3 body & 2-body: $r_\mathrm{cut}=5\mathrm{\AA}$, $n_\mathrm{grid} = 32$ 3-body: $r_\mathrm{cut}=5\mathrm{\AA}$, $n_\mathrm{grid} = 16^3$ & 0.20525 & $O(N_{\rm ranks} t^{(2+3)}_{\rm potential})$ & 136.5\\
      \hline
      ACE-based & $r_\mathrm{cut}=5\mathrm{\AA}$, $n_\mathrm{max}=8$, $l_\mathrm{max}=3$ & 0.21123 & $O(t^\mathrm{(ace)}_\mathrm{potential})$ & 68.4\\
    \hline
    \end{tabular}
    \caption{Performance test with MGP potentials of SiC. Speed unit: $\mathrm{ms \cdot CPU / (atom \cdot timestep)}$. Force MAE unit: meV/\AA.}
    \label{sup-tab:speed_compare}
\end{table}
\end{center}

2+3 body based mapped GP is an approximation, whose accuracy depends on the grid resolution of the spline interpolation, and the ranks from PCA (for variance).
ACE based mapped GP is exact, which only differs from the original Sparse GP model by floating point errors. To reduce computational cost, we chose to use the $\xi=2$ potential and $\xi=1$ variance, where the $\xi=1$ variance is a good approximation of the $\xi=2$ variance, as shown in the Supplementary Information Figure S\ref{supp-fig:var_corr}.

The comparison of the performance between 2+3 body based MGP force field and the ACE based MGP is shown in the Table \ref{sup-tab:speed_compare}.
The computational cost of the mapped variance of the 2+3 body based model is dominated by the PCA decomposed 3-body variance $O(N_{\rm ranks} t^{(2+3)}_{\rm potential})$.
The computational cost of the mapped force field is $t^{\rm (ace)}_{\rm potential}\sim O(n_{\rm descriptor}^{\xi})$, and that of the mapped variance of ACE based model is $O(n_{\rm descriptor}^{2\xi})$, where the $\xi$ is the power of the kernel. Since we use $\xi=1$ for variance and $\xi=2$ for the force field, the cost of variance is comparable to the force field $O(t^{\rm (ace)}_{\rm potential})$.

\newpage
\section*{Supplementary Figures}

\begin{figure}[H]
    \centering
    \includegraphics[width=\textwidth]{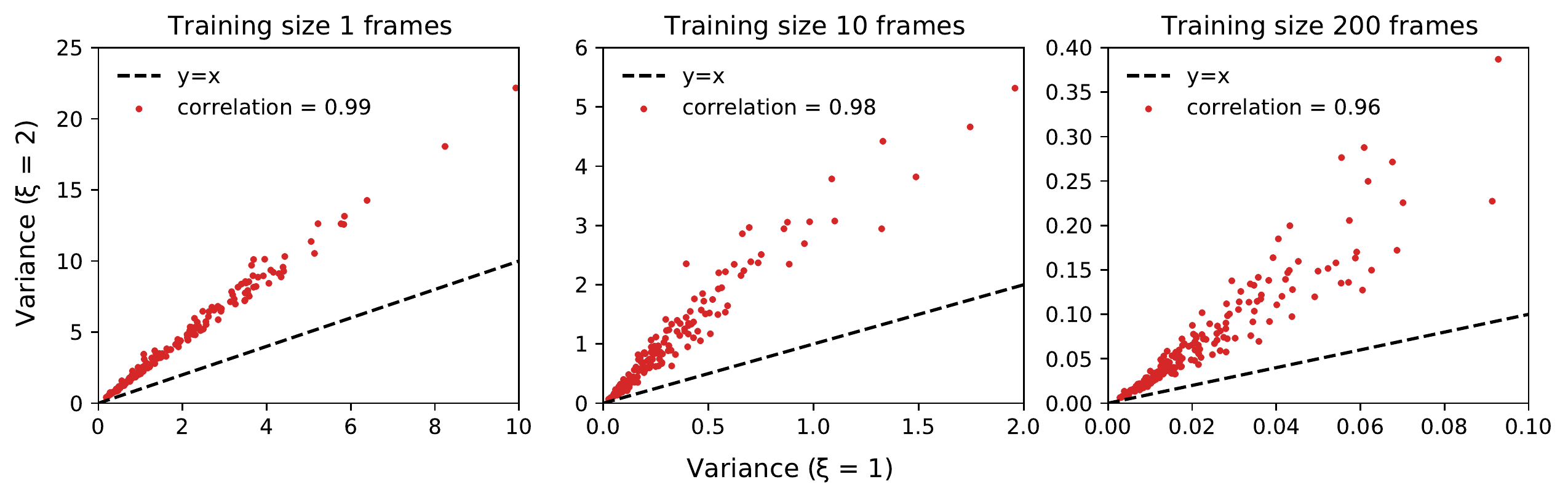}
    \caption{Variance comparison between different powers.}
    \label{supp-fig:var_corr}
\end{figure}

\subsection*{Supplementary Figure \ref{supp-fig:var_corr}: Variance Correlations}

For the computational efficiency, we map the variance with power $\xi=1$ of the normalized inner product kernel of sparse GP for prediction in LAMMPS. While the energy, forces and stress are predicted from power $\xi=2$ kernel, the variance with $\xi=1$ is an approximation of the variance from $\xi=2$ kernel. We test the correlation between the two variances for the same testing configurations with different training set sizes. 
As shown in Fig.S\ref{supp-fig:var_corr}, the two variances have a high correlation above 90\%, even though the values present a constant ratio.
From Supplementary Method, we compare the uncertainty of testing data with the average training set uncertainty in the active learning. Therefore, the constant ratio does not make much difference as long as the approximate uncertainty preserves the order of high/low uncertainties.

\begin{figure}[H]
    \centering
    \includegraphics[width=\textwidth]{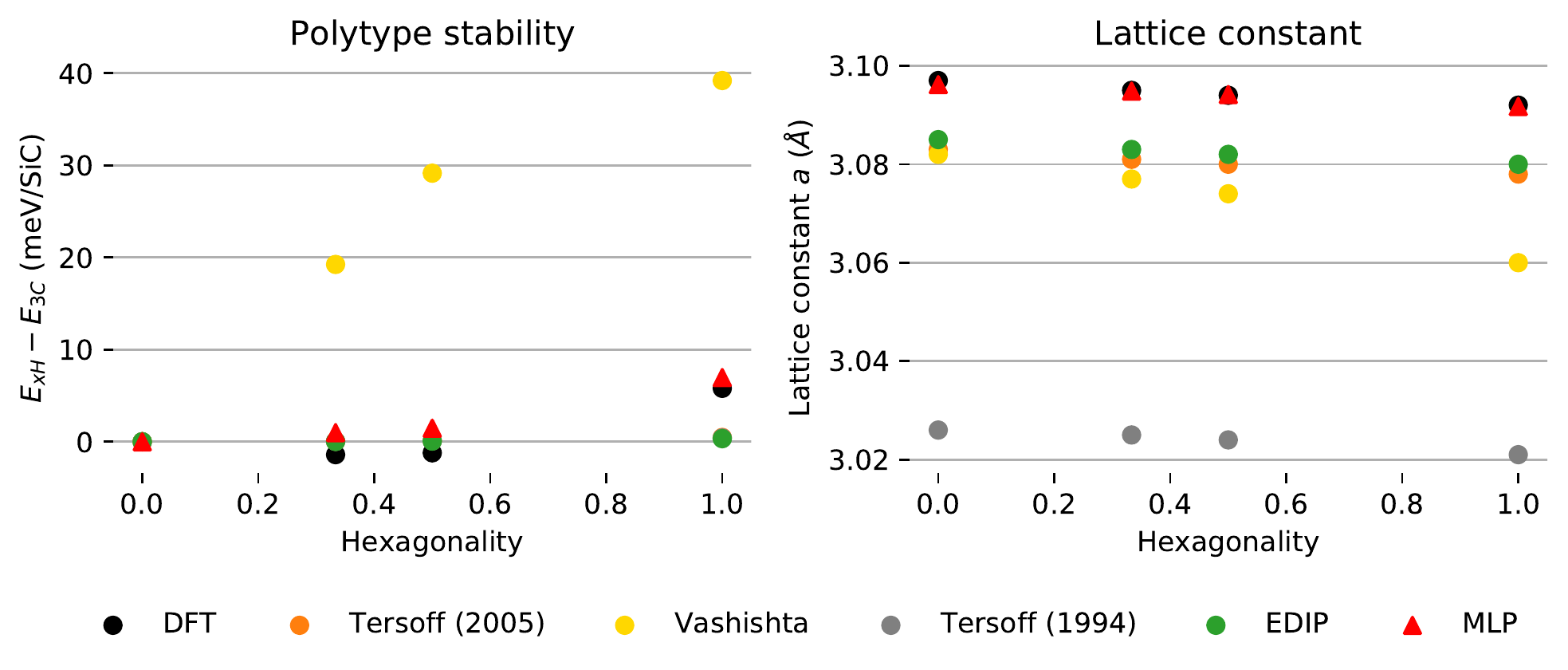}
    \caption{Polytype stability and lattice constant of 3C (hexagonality=0), 6H (hexagonality=1/3), 4H (hexagonality=1/2) and 2H (hexagonality=1), predicted from DFT, FLARE BFF and empirical potentials including Tersoff\cite{erhart_analytical_2005,tersoff_chemical_1994}, Vashishta\cite{vashishta_interaction_2007} and EDIP\cite{jiang_carbon_2012}.}
    \label{supp-fig:poly-en}
\end{figure}

\newpage
\subsection*{Supplementary Figure \ref{supp-fig:poly-en}: Polytype Stability}

We compare FLARE BFF with DFT and empirical potentials on the polytype stability and lattice constant at zero pressure. 
Specifically, we compute ground state energy per Si-C pair and compare the energy differences between hexagonal polytypes (2H, 4H and 6H) with the cubic one (3C). 
While the DFT energy differences between 4H, 6H and 3C are as small as $\sim 1$ meV, 2H is significantly more unstable than the other polytypes by $\sim 6$ meV. 
As shown in Fig.S\ref{supp-fig:poly-en}a, empirical methods such as Tersoff\cite{tersoff_chemical_1994,erhart_analytical_2005} and EDIP\cite{jiang_carbon_2012} do not distinguish the energy differences between different polytypes, and Vashishta\cite{vashishta_interaction_2007} potential predicts very large differences which over-stabilizes the cubic type, while the FLARE BFF reaches around 1 meV error with DFT and gives the correct polytype stability. 
The lattice constants of different polytypes are also compared in Fig.S\ref{supp-fig:poly-en}b, where the discrepancy between FLARE BFF and DFT is around 0.002\AA. And all methods manage to predict the decreasing trend while the hexagonality increases.

\begin{figure}[H]
    \centering
    \includegraphics[width=\textwidth]{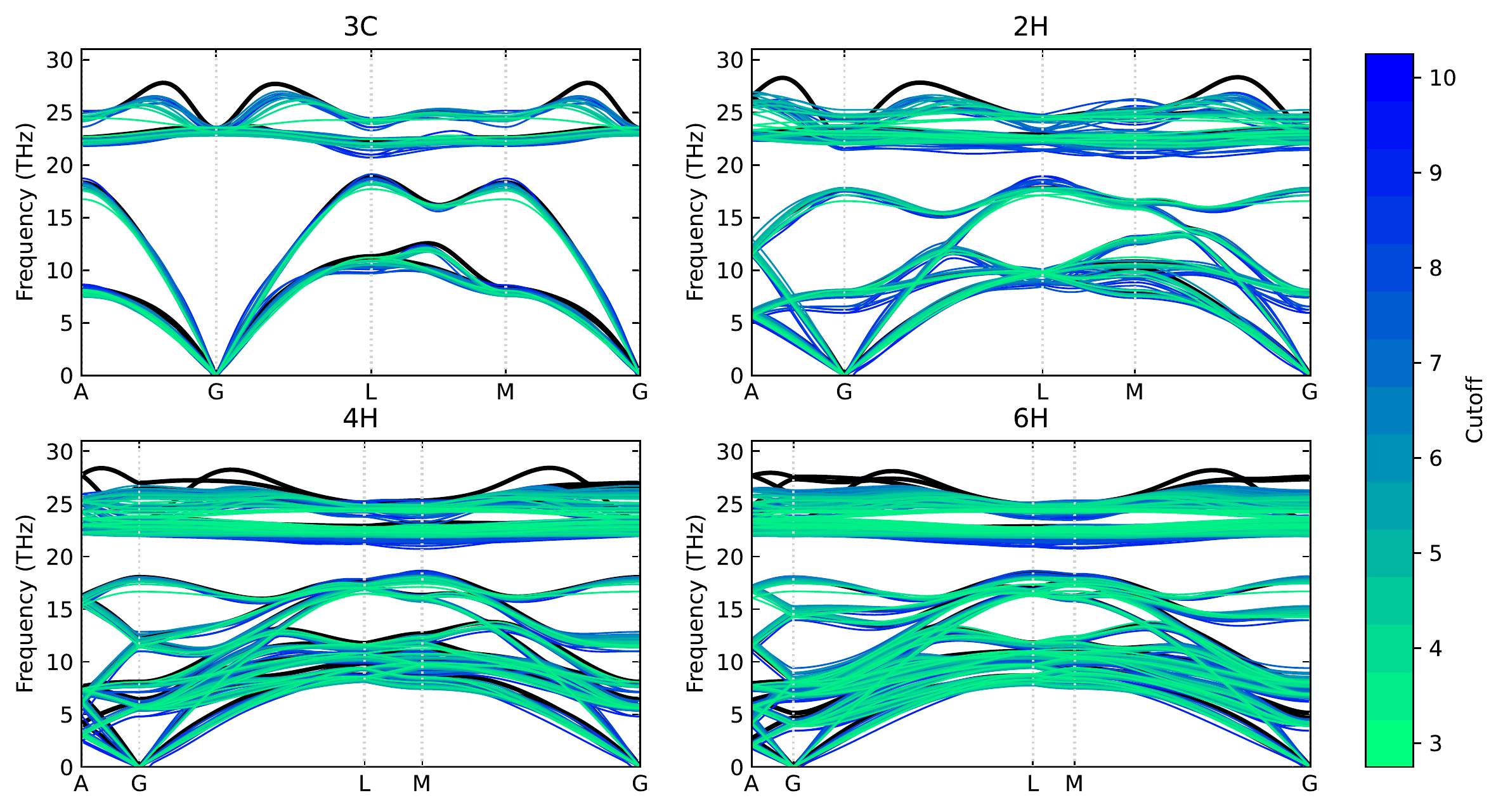}
    \caption{}
    \label{supp-fig:phonon}
\end{figure}

\newpage
\subsection*{Supplementary Figure \ref{supp-fig:phonon}: Phonon Dispersion from Models with Different Cutoffs}

As shown in Fig.5 of the main text, the FLARE BFF and DFT phonon dispersions have an excellent agreement in the acoustic and transverse optical (TO) branches, but show mismatch in the longitudinal optical (LO) branches. 
Considering the LO-TO splitting, even though we do not include non-analytical correction, it is still likely that the polarization makes a non-trivial contribution and is not captured due to the near-sightedness of our machine learning model. 
Therefore, we tested how the model cutoff affects the phonon. 
We trained the model with cutoff from 3 to 10\AA, and the phonons predicted by the models are plotted from light to dark colors. 
As shown in Fig.S\ref{supp-fig:phonon}, all models have an excellent agreement with DFT in acoustic and TO branches, even with a cutoff as short as 3\AA. 
While in the LO branches, the models with smaller cutoffs predict significantly flatter bands than the ones with larger cutoffs.
Though with the largest cutoff our FLARE BFF still does not predict the LO branches perfectly, this result shows an evidence that one source of the mismatch in the LO branches is the long-range effect.
Further investigation is needed such as introducing long-range interactions to the FLARE BFF, and will be left to our future work.

\begin{figure}[H]
    \centering
    \includegraphics[width=0.8\textwidth]{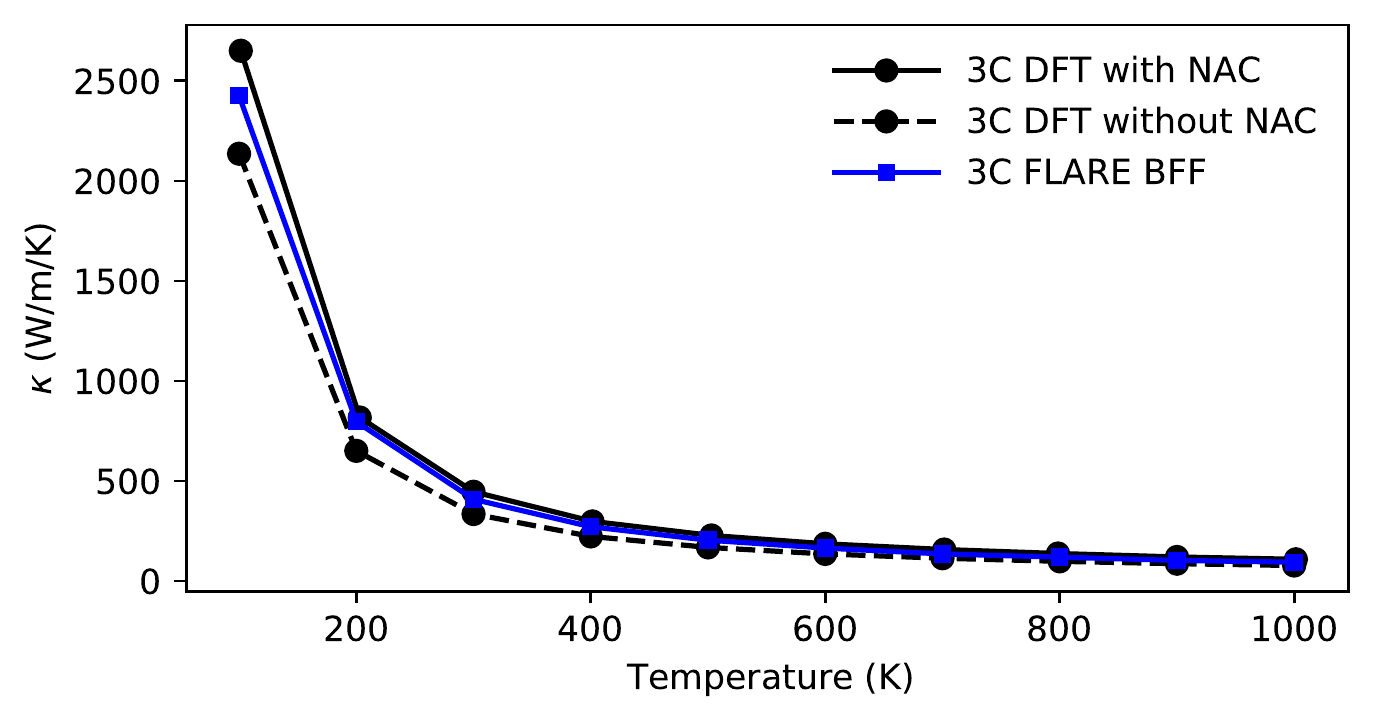}
    \caption{Thermal conductivity without and with NAC \cite{protik_phonon_2017} at different temperatures.}
    \label{supp-fig:nac}
\end{figure}

\newpage
\subsection*{Supplementary Figure \ref{supp-fig:nac}: Non Analytic Correction on the Thermal Conductivity}
We calculated the thermal conductivity from Boltzmann transport equation with the non analytical correction (NAC) term from Born effective charge, and compared it to that without NAC term at temperatures from 100 to 1000 K.
Fig.S\ref{supp-fig:nac} indicates that the exclusion of the non analytical correction term underestimates the thermal conductivity, but the underestimation is not significant.
Since FLARE BFF is trained on the DFT data without long-range electrostatic interactions, and the force constants are computed without NAC term, it underestimates thermal conductivity compared to the DFT with NAC as expected.

\begin{figure}[htbp]
    \centering
    \includegraphics[width=0.9\textwidth]{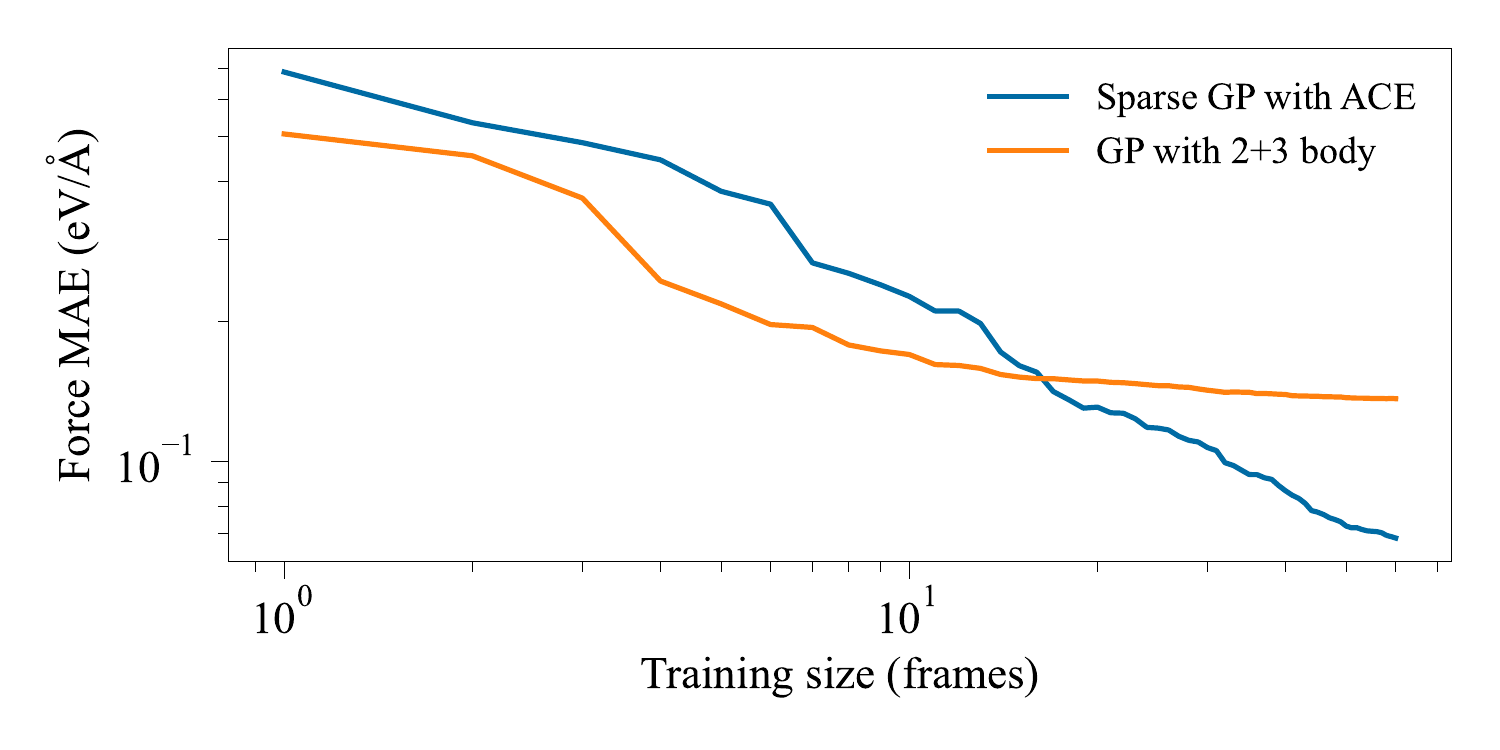}
    \caption{Comparison of accuracy between ACE-based and 2+3 body based Gaussian process force field on the bulk SiC dataset.}
    \label{sup-fig:comparison}
\end{figure}

\newpage
\subsection*{Supplementary Figure \ref{sup-fig:comparison}: Accuracy Comparison of 2+3-body and ACE-based Models}

A comparison of accuracy between the GP model with 2+3 body descriptors and the one with ACE descriptors is made with the SiC bulk dataset, as shown in Fig.S\ref{sup-fig:comparison}.

\newpage
\bibliographystyle{naturemag}
\bibliography{main}